# Damping effects of viscous dissipation on growth of symmetric instability

Laur Ferris[a], Donglai Gong[a]

[a]*Virginia Institute of Marine Science - William & Mary, Gloucester Point, VA 23062, USA*

Symmetric instability (SI) is a frontal instability arising from the interaction of rotation with lateral and vertical shear of a frontal jet and is a generalization of shear, centrifugal, and gravitational instabilities. While the onset of SI has been studied in numerous observations and models, intuition about its growth in physical ocean comes primarily from constant-viscosity linear instability analysis and large eddy simulation (LES). A forward cascade arising from SI in the real ocean, where numerous fine-to-microscale processes interact with growing SI velocity cells, is less understood. While many instances of symmetrically unstable flow have been observed, observations of enhanced turbulent kinetic energy (TKE) dissipation ($\epsilon$) at these sites are less common. We use numerical instability analysis of an idealized geostrophic jet to show that viscous-diffusive effects of preexisting turbulence from other turbulent processes (e.g., competing instabilities, internal wave processes, or boundary layer processes) can suppress the growth of SI in the real ocean. For example, a moderate level of ambient turbulence, represented by uniform diffusivity and viscosity of $\kappa = \nu = 10^{-4}$ m²/s, restricts the wavelength range of SI's fastest-growing mode from $\mathcal{O}$(10-100)m to $\mathcal{O}$(100)m and elongates its e-folding timescale by $\mathcal{O}$(1-10) hrs; suggesting the net viscous-diffusive effects of preexisting turbulence can damp the growth of SI. Viscous damping is one possible explanation for the rarity of SI structures in the real ocean, and our results motivate the inclusion of dependence on previous-timestep $\epsilon$ or $\kappa$ when parameterizing SI in regional models.

*Email address for correspondence: lnferris@vims.edu*

---

**Introduction**

There are several physical processes besides shear-convective instability, waves, and Langmuir circulation that can support mixing in the surface boundary layer and main thermocline but are not widely parameterized in Reynolds Averaged Navier-Stokes (RANS)-type circulation models. One class of processes are submesoscale frontal instabilities, which can extract TKE for mixing from a combination of baroclinicity and shear at fronts, often perpetuated by surface forcing. These include SI, as well as the more general mixed layer instability[1] (MLI) events described by Callies et al., 2016. SI has wavelength 20-500 m (Dong et al., 2021a) such that it is unresolved in most ocean models. The issue of including submesoscale frontal instabilities in models (either explicitly or through subgrid-scale parameterizations) is an area of active research, and most subgrid-scale parameterizations of turbulence-generating processes leverage vertical gradients of buoyancy and velocity. At the same time, SI is strongly dependent on lateral gradients and thus is unrepresented in most model diffusivity closures.

---

[1]MLI is a terminology for eddies formed by ageostrophic baroclinic instability (ABI) in a surface mixed layer, having weak stratification and a small $\mathcal{O}$(0.1) Prandtl ratio, with spatial scales $\mathcal{O}$(1-10km) and growth timescales $\mathcal{O}$(1 day) (Boccaletti et al., 2007). ABI arises from the same setup as SI but acts in the along-front direction, generating strong restratification by extracting available potential energy. ABI can occur at the submesoscale or act at the mesoscale.

There are ongoing community efforts to incorporate SI into ocean models. Dong et al. (2021b) applied the geostrophic-and-forcing-dependent Bachman et al. (2017) parameterization for mixed layer SI to Coastal and Regional Ocean Community Model (CROCO)-ROMS and found that its inclusion yielded energy extraction similar to an SI-resolving LES. Yankovsky et al. (2021) have developed a parameterization for SI which does not rely on dimensional parameters as does that of Bachman et al. (2017). Critical to implementing these parameterizations is the study of their usage; with increasing model resolution, parameterizations must be used with caution. Bachman & Taylor (2014) examine the issue of partially-resolved SI and double-counting (where mixing effects of resolved SI are duplicated by SI mixing parameterizations), using a 2D MITgcm model with 5-m vertical resolution to show that large horizontal viscosity (80 m$^2$/s) paired with a fine grid (250-3000m) can prevent mixed layer restratification, while smaller horizontal viscosity (10 m$^2$/s) has been demonstrated to cause excessive restratification due to nonphysical mixing of numerical origin. In effect, KPP and other subgrid scale parameterization schemes prescribing large mixed layer eddy viscosities can prevent growth of SI modes, called *grid-arrested re-stratification* (Bachman & Taylor, 2014). While non-physical grid arrested stratification is an insidious issue, we are interested in the physical damping effect of background turbulence on growth of SI.

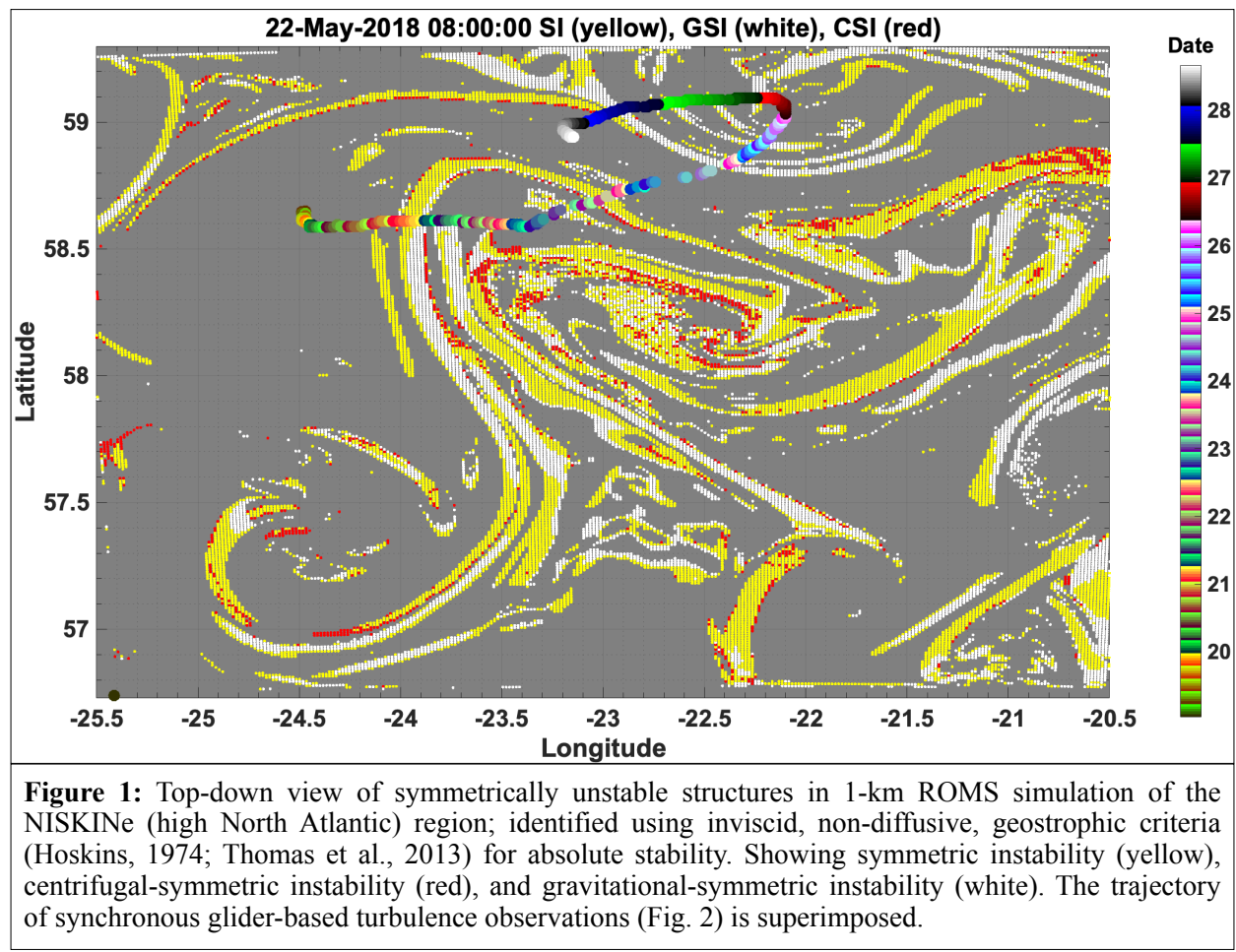

**Figure 1:** Top-down view of symmetrically unstable structures in 1-km ROMS simulation of the NISKINe (high North Atlantic) region; identified using inviscid, non-diffusive, geostrophic criteria (Hoskins, 1974; Thomas et al., 2013) for absolute stability. Showing symmetric instability (yellow), centrifugal-symmetric instability (red), and gravitational-symmetric instability (white). The trajectory of synchronous glider-based turbulence observations (Fig. 2) is superimposed.

Recent LES studies have suggested that pure symmetric instability is generally difficult to realize, achieved only though no wind stress and weak cooling (Skyllingstad & Samelson, 2020). Skyllingstad et al. (2017) discuss the innate dependence of eddy viscosity on geostrophic shear production and additional geostrophic shear production from wind and wave forcing, illustrating how the evolution of SI is coupled to its boundary conditions. Similarly, Skyllingstad & Samelson (2020) use resolved turbulence in place of constant eddy viscosity to study the evolution of baroclinic instability in the presence of wind forcing. Notably, Wienkers et al. (2021) extend previous numerical solutions of SI using the idealized Eady problem to include viscosity and thus, energy-conscious growth of SI, to show two limiting regimes where buoyancy production and shear production are primarily responsible for TKE in weak and strong fronts, respectively. Whalen et al. (2022; in prep.) point out that most studies have focused on phenomena at individual fronts, and the global geography and characteristics of submesoscale fronts must be quantified in order to understand where different theoretical frameworks may apply.

The addition of viscosity and diffusivity has applications both for the physical representation of growing instabilities, as well as stabilizing the numerical algorithms used to solve the normal mode equations due to singular behavior of the normal mode equations (both Taylor Goldstein and Eady forms) in the vicinity of critical layers (Rees & Monahan, 2014; Zemskova et al., 2020). It is well known that eddy viscosity and diffusivity, the dispersion of momentum and energy due to turbulence processes, suppress stratified shear instability for small Reynolds numbers $Re \equiv UH/\nu < 100$ (Smyth & Carpenter, 2019). As an instability grows, dissipation removes TKE and momentum diffusion redistributes it through the water column; a phenomenon documented by Kaminski & Smyth (2019). However, the effects of varying background turbulence on stratified rotational instabilities are less studied. Notably, Smyth (2008) find that ambient turbulence has a damping effect on the growth of instabilities arising in a baroclinic, double diffusive frontal system; and Smyth & Ruddick (2010) demonstrate its control on the vertical scale of thermohaline intrusions. Smyth et al. (2012) test a second-moment closure for mixing based on these findings, though the authors did not find the closure significantly improved upon an empirical model.

In this study we conduct numerical linear instability analysis on an idealized geostrophic jet, varying its stratification and shear to represent various SI flows in the stratified ocean surface boundary layer and main thermocline, to show that background turbulence reduces both the growth rate ($\sigma$) and range of unstable wavenumbers ($k, l$). We build the symmetrically unstable jet using the observed Prandtl ratio and Richardson numbers ($P = f/\sqrt{B_z}, Ri = B_z/U_z^2$) present in 1-km ROMS simulations of the Icelandic Basin of the high North Atlantic (Fig. 1) and Antarctic Circumpolar Current, developed during the Office of Naval Research (ONR)'s

4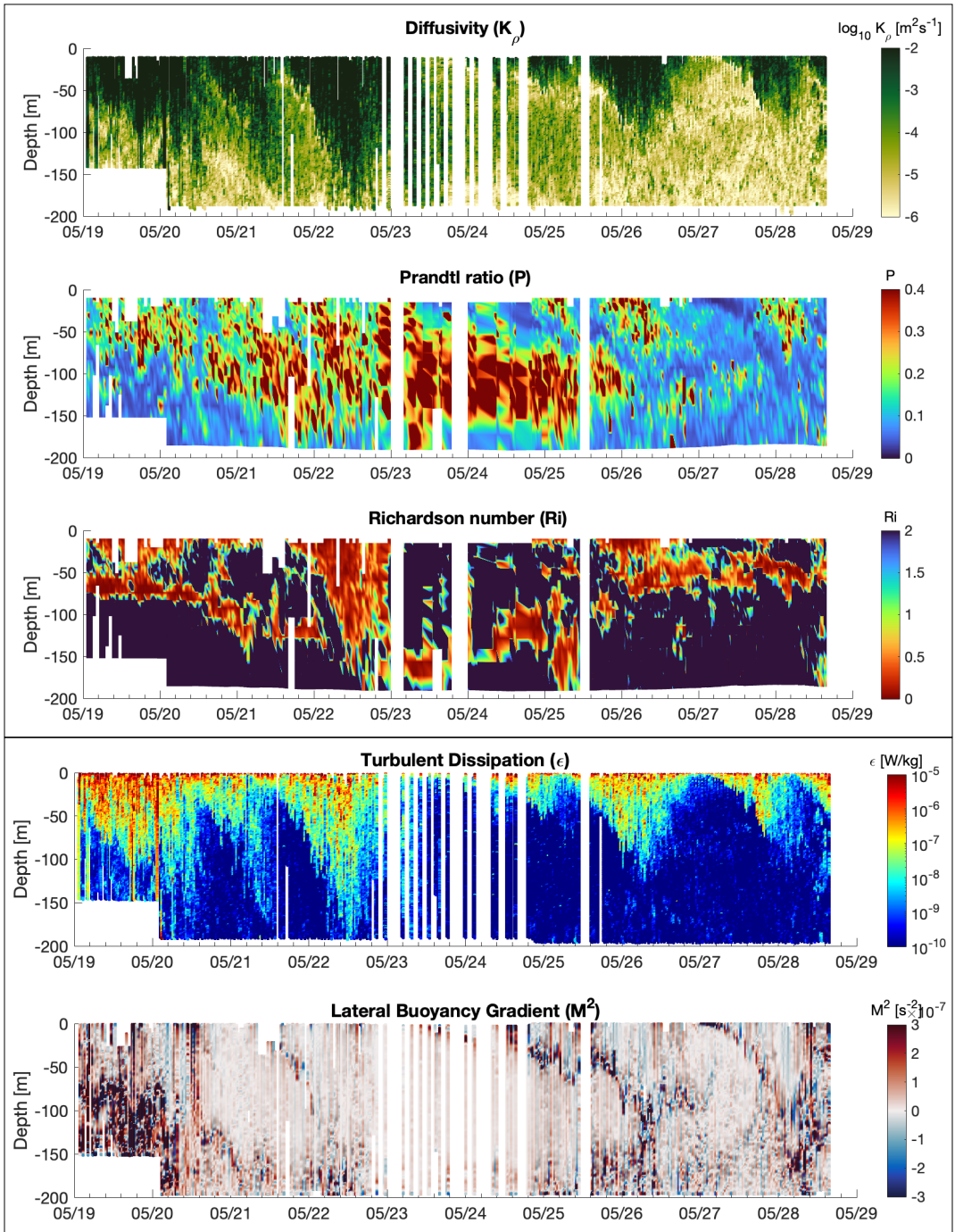

**Figure 2:** Observed and derived variables for NISKINe. Showing (a) estimated diffusivity from the Osborn (1980) relation applied to glider microstructure and CTD, **(b)** Prandtl ratio $P = f/\sqrt{B_z}$ and **(c)** Richardson number $Ri = B_z/U_z^2$ derived from ROMS velocity and glider CTD, **(d)** direct measurements of TKE dissipation rate, and **(e)** lateral buoyancy gradient from glider CTD. On 22-May-2018 and 26-May-2018, TKE dissipation rate significantly exceeded levels predicted by using boundary layer similarity scaling for shear-convective turbulence, and Richardson numbers $0.25 < Ri \lesssim 1$ suggest a plausible range for SI. The glider crossed into the warm side of the front on 20 May 2018 and smaller filaments during 26-29 May 2018.

NISKINe DRI (Near Inertial Shear and Kinetic Energy in the North Atlantic Departmental Research Initiative) and the AUSSOM (Autonomous Sampling of Southern Ocean Mixing) project; as well as glider observations (Fig. 2) from each region[2]. We find that viscous-diffusive effects of background turbulence constrain the length scales of SI overturning cells and slows their development, providing greater opportunity for secondary interactions with internal waves and other physical processes in the real ocean.

**Methods**

In general, linear instability analysis (analytical or numerical) is conducted by linearizing the equations of motion of the fluid system by imposing a small perturbation, deriving an ordinary differential equation (ODE) governing the flow assuming a wave-like solution, and examining the integral properties of the ODE to find where the wave-like solution takes on an exponential growth rate $\sigma = \sigma_r + i\sigma_i$ such that $\sigma_r > 0$ corresponds to a growing instability. Two examples of ODEs are the Rayleigh equation (for a parallel shear flow) and the Taylor-Goldstein equation (for a stratified shear flow). A nonzero imaginary component $\sigma_i$ means the wave-like solution is oscillatory. Equivalently $\{\sigma\}_i = -\tilde{k}c$ such that an instability has a complex phase speed. The growth rate is related to the e-folding time (for a disturbance to grow by a factor of $e \approx 2.72$) by $\sigma_r^{-1}$. Instability analysis can be conducted on real flows through numerical solution of the normal mode equations (using singular value decomposition, a spectral method such as Fourier-Galerkin, or recast as an eigenvalue problem as done below).

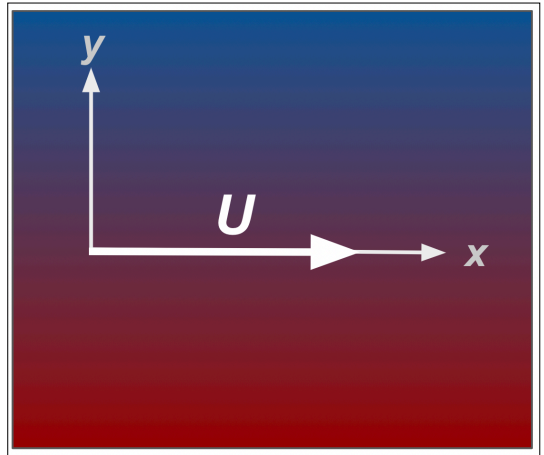

**Figure 3:** Top-down diagram of a frontal jet. Red indicates greater buoyancy and $U$ represents the mean flow.

We begin with a stratified, rotating system, after Smyth & Carpenter (2019), but with the addition of background viscosity and diffusivity, as well as lateral shear. After Smyth & Carpenter (2019), we start with the linearized governing equations for a perturbation in a rotating, stratified geostrophic jet $\vec{U} = [U(y,z); 0; 0]$ (Fig. 3) and add terms for viscosity ($\nu$) and diffusivity ($\kappa$):

$$\vec{\nabla} \cdot \vec{u} = 0 \qquad (1a)$$

$$\left(\frac{\partial}{\partial t} + U\frac{\partial}{\partial x}\right)\vec{u} + vU_y \hat{e}^{(x)} + wU_z \hat{e}^{(x)} = -\vec{\nabla}\pi + b\hat{e}^{(z)} + \vec{u} \times f\hat{e}^{(z)} + \nu\nabla^2 \vec{u} \qquad (1b)$$

---

[2] For further details about NISKINe, the reader is referred to the *Oceanography Special Issue: Near-Inertial Shear and Kinetic Energy in the North Atlantic Experiment (NISKINe)*. For further details about AUSSOM, the reader is referred to Ferris (2022) and Ferris et al. (2022a; 2022b).

$$\left(\frac{\partial}{\partial t} + U\frac{\partial}{\partial x}\right)b + vB_y + wB_z = \kappa\nabla^2 b \tag{1c}$$

where lowercase variables are perturbation quantities. We make several simplifying assumptions about the viscosity and diffusivity, a treatment that is by no means comprehensive but aims to demonstrate the point that dispersion by preexisting turbulence can damp the growth of SI: (1) There exists a scale separation such that the three scales of motion are that of the submesoscale flow ($U$), the perturbation ($u$), and that of unresolved microscale turbulence ($u'$) below the vertical grid scale, $\leq 1$ m. In other words, while damping of the system is nonlinear, the viscous term is constant such that and $\kappa = \nu$ does not depend on the dynamics of the growing instability. (2) Dispersion of momentum and energy by this preexisting microscale turbulence has an associated eddy diffusivity and viscosity that matches or exceeds the molecular viscosity such that $\nu\nabla^2 u + \nu_{mol}\nabla^2 u \approx \nu\nabla^2 u$. (3) Using an isotropic eddy diffusivity and viscosity applies a conservative level of dispersion; in reality horizontal diffusivity is equivalent to or higher than diapycnal diffusivity (following the dimensional arguments of Liu et al., 2012).

There are two ways to proceed: One setup admits both symmetric and baroclinic instability (Eq. 1 with normal mode solution $\phi' = \hat{\phi}(z)e^{\sigma t + i(kx+ly)}$ such that necessarily $U_y \equiv 0$). The alternative setup admits symmetric and centrifugal instability by assuming no along-front variation in the base state or perturbation (such that $\partial\xi/\partial x \equiv 0$, where $\xi = u, v, w, b$). We employ the former such that coefficients of perturbation quantities can only depend on $z$, $U_y \equiv 0$, and $B_y$ is constant[3] over the scale of solution.

We refer the reader to Section 8.8 of Smyth & Carpenter (2019); as this problem is similar but with the addition of terms for viscosity and diffusivity. After algebra (Appendix A), and we obtain a reduced system of equations:

$$\sigma\begin{pmatrix} I & 0 & 0 \\ 0 & \nabla^2 & 0 \\ 0 & 0 & I \end{pmatrix}\begin{pmatrix} \hat{q} \\ \hat{w} \\ \hat{b} \end{pmatrix} = \begin{pmatrix} -ikU + \nu\nabla^2 & ilU_z + fD^{(1w)} & 0 \\ -fD^{(1q)} & -ikU\nabla^2 + \nu\nabla^4 & -\tilde{k}^2 \\ \frac{ik}{\tilde{k}^2}B_y & -\frac{il}{\tilde{k}^2}B_y D^{(1w)} - B_z & -ikU + \kappa\nabla^2 \end{pmatrix}\begin{pmatrix} \hat{q} \\ \hat{w} \\ \hat{b} \end{pmatrix} \tag{2a}$$

$$\hat{u} = \hat{v}k/l + \hat{q}i/l \tag{2b}$$
$$\hat{v} = i(\hat{w}_z l - \hat{q}k)/\tilde{k}^2 \tag{2c}$$
$$\hat{\pi} = -(\sigma\hat{v} + ikU\hat{v} + f\hat{u})/(il) + \nu\nabla^2\hat{v}/(il) \tag{2d}$$
$$\hat{\eta} = w/(\sigma + ikU) \tag{2e}$$

---

[3] The chosen normal mode solution $\phi' = \hat{\phi}(z)e^{\sigma t+i(kx+ly)}$ requires that coefficients can only depend on $z$. $U_y \equiv 0$ because the second terms of (Eq. 1b) and (Eq. 1c) contain coefficient $U\partial/\partial x$. Stratification similarly cannot vary in $y$ such that $B_{zy} = B_{yz} = 0$. Thus $B_y$ is constant and the differentiated thermal wind $B_{yz} = -fU_{zz}$ implies $U_{zz} \equiv 0$.

where $D^{()}$ are vertical derivative operators[4], $\sigma$ are eigenvalues for the growth rate, and $\hat{q}, \hat{w}, \hat{b}, \hat{u}, \hat{v}, \hat{\pi}, \hat{\eta}$ are eigenfunctions for the perturbation quantities. This system is analogous to the Taylor-Goldstein equation with the effects of rotation, viscosity, and diffusivity added.

The problem is nondimensionalized using a time scale $T \equiv 1/U_z$ and length scale $H$, with flow characteristics defined by the Prandtl number ($Pr$), Reynolds number ($Re$), and Richardson number ($Ri$), and Prandtl ratio ($P$) (Table 1). Wienkers et al. (2021) more elegantly used front

| Table 1. Variables and dimensionless [*] forms | | |
|---|---|---|
| Prandtl ratio | [ ] | $P = f/\sqrt{B_z}$ |
| Prandtl number | [ ] | $Pr = \nu/\kappa$ |
| Richardson number | [ ] | $Ri = B_z/U_z^2$ |
| Reynolds number | [ ] | $Re = U_z H^2/\nu$ |
| domain height | [m] | $H^* = H/H$ |
| vertical coordinates | [m] | $z^* = [-H^*/2, H^*/2]$ |
| vertical shear | [s⁻¹] | $U_z^* = U_z/U_z = U_z T$ |
| velocity | [ms⁻¹] | $\vec{u^*}, U^* = U/U_z H$ |
| time | [s] | $t^* = t U_z$ |
| spatial derivative | [m] | $\vec{\nabla^*} = H\vec{\nabla}$ |
| stratification | [s⁻²] | $B_z^* = B_z/U_z^2 = Ri$ |
| buoyancy | [ms⁻²] | $b^*, B^* = B/H U_z^2$ |
| pressure/density | [m²s⁻²] | $\pi^* = \pi/U_z^2 H^2$ |
| Coriolis parameter | [s⁻¹] | $f^* = f/U_z = P\sqrt{Ri}$ |
| kinematic viscosity | [m²s⁻¹] | $\nu^* = \nu/U_z H^2 = 1/Re$ |
| diffusivity of density | [m²s⁻¹] | $\kappa^* = \nu^*/Pr$ |
| along-front wavenumber | [m⁻¹] | $k^* = kH \approx [0,4]P$ |
| across-front wavenumber | [m⁻¹] | $l^* = lH \approx [1, \pi/\Delta z^*]P$ |
| growth rate | [s⁻¹] | $\sigma^* = \sigma T = \sigma/U_z$ |

---

[4] Operators $D^{()}$ are matrices which numerically differentiate the multiplicand to the order specified in parenthesis, with boundary conditions appropriate for the indicated variable. For example $D^{(1q)}$ applies $d/dz$ with boundary conditions appropriate for multiplicand $q$.

strength $\Gamma = M^2/f^2$ (where $M^2$ is the lateral buoyancy gradient) as the fourth dimensionless number, but we use the Prandtl ratio ($P$) to describe the amount of stratification ($B_z$) for a given latitude on earth ($f$). From a practical perspective, those aiming to compare the key result of this paper (Fig. 9) to observations at geostrophic jets will not require $B_y$ (which can be expensive to measure).

We use fixed boundary conditions for velocity, frictionless boundary conditions for vorticity, and insulating boundary conditions for buoyancy. The lateral buoyancy gradient is derived from thermal wind balance, $B_y = -fU_z$. We use $U_z = 5 \times 10^{-4}$ s⁻¹ and $H = 100$ m to re-dimensionalize quantities following numerical solution, representing a typical submesoscale current in the upper ocean. This derivation assumes isotropic eddy viscosity ($\nu$) and diffusivity ($\kappa$) such that $Pr = 1$[5], which are both uniform with depth. The viscous-diffusive effects of background turbulence are varied from $\kappa = 10^{-6}$ to $10^{-2}$ m²/s based on estimated diffusivity from glider microstructure measurements (Fig. 4) in the AUSSOM and NISKINe regions. This method is not novel; Liu et al. (2012) similarly chose viscosity and diffusivity profiles from the TKE dissipation rate by leveraging the Osborn (1980) relation and assuming $Pr \approx 1$. The eddy viscosity is used to prescribe $Re$, similar to Lian et al. (2020), along with $P$ and $Ri$ chosen from 1-km ROMS simulations. We use $NZ = 100$, 1-m vertical levels, which dictates the highest resolved unstable wavenumbers.

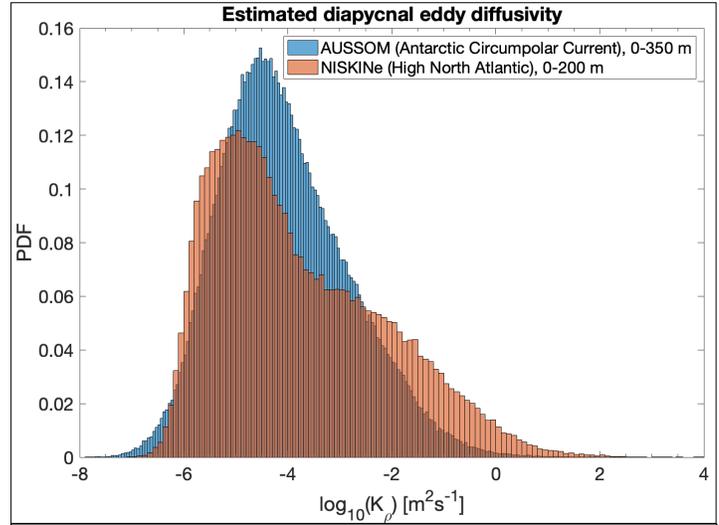

**Fig. 4:** Diapycnal eddy diffusivity estimated from the Osborn (1980) relation, with $\kappa_\rho = \Gamma \epsilon / N^2$ and $\Gamma = 0.2$, from microstructure measurements in each study region. Smyth (2020) discuss the physical explanation for and limitations of this approximation.

The scaled continuity equation (Eq. 3a) is obtained by substituting dimensional-to-dimensionless equivalencies into (Eq. 1a) and dividing through by $U_z$. The scaled momentum equation (Eq. 3b) is obtained by substituting dimensional-to-dimensionless equivalencies into (Eq. 1b) and dividing through by $U_z^2 H$. The scaled energy equation (Eq. 3c) is obtained by substituting dimensional-to-dimensionless equivalencies into (Eq. 1c) and dividing by $U_z^3 H$.

$$\overrightarrow{\nabla^*} \cdot \overrightarrow{u^*} = 0 \tag{3a}$$

---

[5] Lian et al. (2020) applies anisotropic horizontal and vertical values for $\nu$ and $\kappa$ to the Taylor-Goldstein. We do not know of a published application of anisotropic values in a rotating stratified system.

$$\left(\frac{\partial}{\partial t^*} + U^*\frac{\partial}{\partial x^*}\right)\vec{u^*} + w^*U_z^*\hat{e}^{(x)} = -\vec{\nabla^*}\pi^* + b^*\hat{e}^{(z)} + \vec{u^*} \times P\sqrt{Ri}\hat{e}^{(z)} + \frac{1}{Re}\nabla^{*2}\vec{u^*} \quad (3b)$$

$$\left(\frac{\partial}{\partial t^*} + U^*\frac{\partial}{\partial x^*}\right)b^* + v^*B_y^* + w^*B_z^* = \frac{1}{PrRe}\nabla^{*2}b^* \quad (3c)$$

Diffusion entails quasi-equilibrium such that for linear instability analysis to be valid, the setup must satisfy frozen flow approximation; in other words the instability must grow faster than free diffusion by the scaled kinematic viscosity (Table 1) modifies the mean flow, $\sigma^* \gg 1/Re$. With high Reynolds numbers, this condition is satisfied in most unstable cases (Fig. 4) and results are disregarded when it is unsatisfied.

After Smyth & Carpenter (2019), the budget for evolution of perturbation kinetic energy ($K_t$) is obtained by dotting the momentum equation (Eq. 1b) with velocity vector $\vec{u'}$; and using the equivalence between perturbation ($a'$) and normal mode quantities ($\hat{a}$), $\overline{a'b'} = \frac{1}{2}\{\hat{a}\hat{b}^*\}_r e^{2\sigma t}$ with $t = 0$ and $r$ denoting the real part.

$$K_t = 2\sigma_r K = SPZ + SPY - EF_z + BF + \nu K_{zz} - \epsilon \quad (4a)$$
$$K = (|\hat{u}|^2 + |\hat{v}|^2 + |\hat{w}|^2)/4 \quad (4b)$$
$$SPZ = -U_z\{\hat{u}^*\hat{w}\}_r/2 \quad (4c)$$
$$SPY = -U_y\{\hat{u}^*\hat{v}\}_r/2 \quad (4d)$$
$$EF = \{\hat{w}^*\hat{\pi}\}_r/2 \quad (4e)$$
$$BF = \{\hat{w}^*\hat{b}\}_r/2 \quad (4f)$$
$$\epsilon = \nu(|\hat{u}_z|^2 + |\hat{v}_z|^2 + |\hat{w}_z|^2 + 4\tilde{k}^2 K)/2 \quad (4g)$$

Here $K$ is the perturbation kinetic energy, $SPZ$ is vertical shear production, $SPY$ is lateral shear production (zero for this setup), $EF_z$ is the divergence of pressure-driven vertical kinetic energy flux, $BF$ is the buoyancy production, and $\epsilon$ remains viscous dissipation of TKE [W/kg]. For (Eq. 4) only, "*" indicates a complex conjugate (for all other instances "*" indicates a dimensionless variable). In general, shear production is dominant in CI and SI and buoyancy production is

| Table 2. Dimensionless numbers representative of symmetrically unstable oceanic flows | | |
|---|---|---|
| **Regime** | $Ri$ | $P$ |
| A: NISKINe surface boundary layer | 0.3 | 0.1 |
| B: AUSSOM surface boundary layer | 0.5 | 0.2 |
| C: NISKINe or AUSSOM shallow thermocline | 0.95 | 0.05 |

dominant in baroclinic and gravitational instability. Both $EF_z$ and $\nu K_{zz}$ spread but do not create energy (by pressure and viscosity, respectively), causing accumulation at a particular depth. The purpose of formulating a TKE budget is to validate (Eq. 2a) and its solution algorithm, i.e. a centrifugal-symmetric instability should indeed be characterized by shear production. The TKE budget is reasonable for a flow with uniform shear and stratification; with predominately shear production as expected for pure SI, pressure-driven flux ($EF$) vanishing at the boundaries.

Dimensionless numbers representative of typical symmetrically unstable flows are obtained from the ROMS hindcasts for the NISKINe and AUSSOM regions (Table 2), organized by increasing stratification ($P$). The surface boundary layer in NISKINe is more weakly stratified than in AUSSOM; with higher maximum vertical shear in the latter. Both regions are at similar latitude and thus comparable planetary control. We examine principally the fastest-growing mode, and color axes (Fig. 5, representing growth rates) are scaled using the maximum analytical growth rates[6].

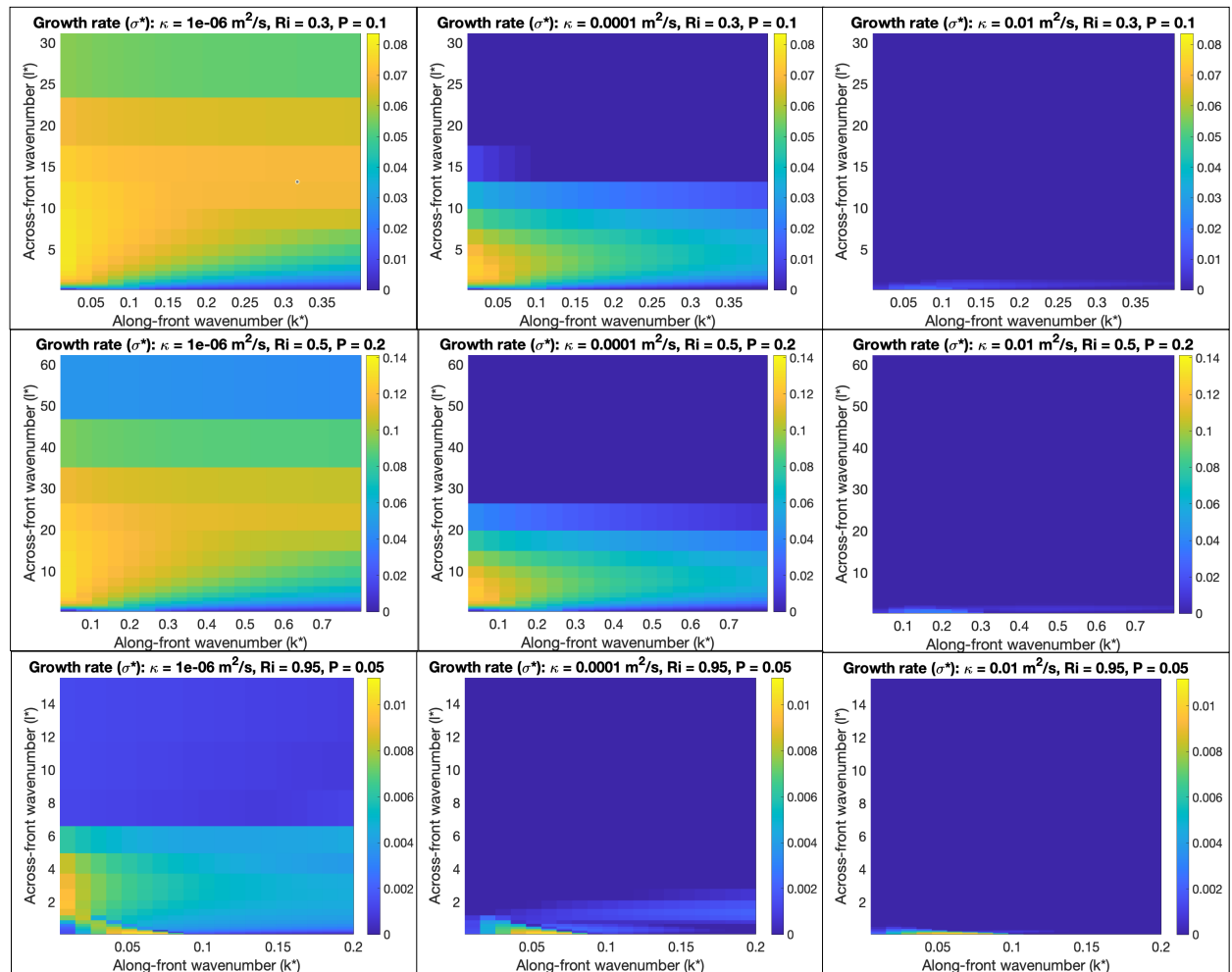

**Figure 5:** Dimensionless growth rates ($\sigma^*$) as a function of along-front wavenumber ($k^*$) and across-front wavenumber ($l^*$) for increasing levels of ambient diffusivity, $\kappa = 10^{-6}$, $\kappa = 10^{-4}$, and $\kappa = 10^{-2}$ m²/s; for which growth rates below $\sigma^* \gg 1/Re = 2 \times 10^{-7} - 10^{-3}$ (dark blue) are not meaningful. Each row corresponds to flow described in Table 2. The symmetric mode is found along the *y*-axis, whereas the baroclinic mode is found along the *x*-axis. Growth of SI in the mixed layer cases (top and center rows) is more resilient to eddy viscous-diffusive effects than in the more stratified thermocline, where the baroclinic mode is relatively undamped by turbulent effects.

**Results**

The diffusivity levels and Reynolds numbers in this study are comparable to previous fixed-viscosity models. Our setup corresponds to varying $Re = 5 \times 10^2 - 10^6$; for comparison Wienkers et al. (2021) used $Re = 10^5$, and Stamper & Taylor (2017) used a fixed $\kappa = \nu = 6 \times 10^{-4}$ m²/s in Diablo-based numerical simulations. Both studies, which used low-to-moderate effective turbulence levels, similarly achieved growing SI but with different motivations (we study the effect of turbulence on spatial scale and timescale of SI). Solving (Eq.

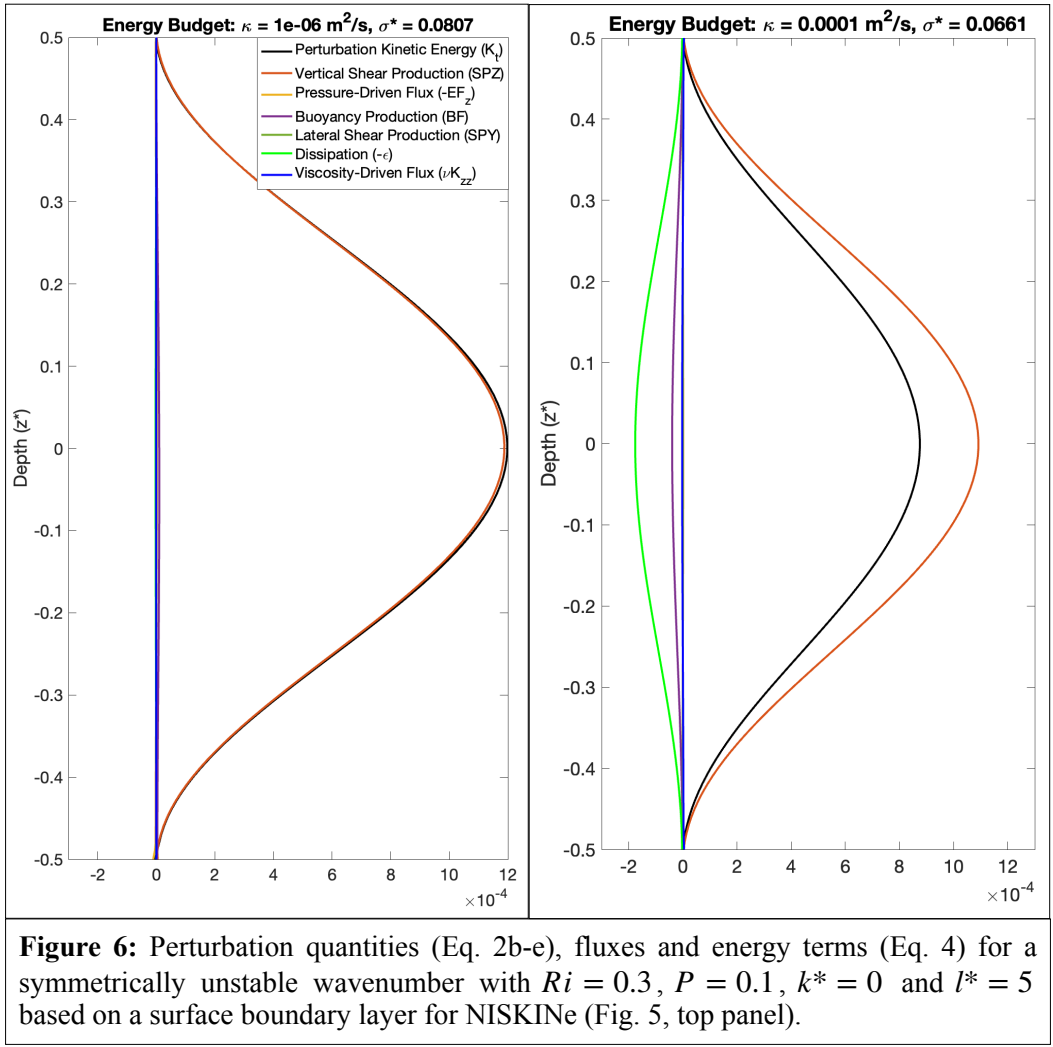

**Figure 6:** Perturbation quantities (Eq. 2b-e), fluxes and energy terms (Eq. 4) for a symmetrically unstable wavenumber with $Ri = 0.3$, $P = 0.1$, $k^* = 0$ and $l^* = 5$ based on a surface boundary layer for NISKINe (Fig. 5, top panel).

---

[6] Analytical maximum growth rates for the symmetric, centrifugal, and Eady modes are $\sigma_{SI} = |f|(1/Ri - \zeta_a/f)^{1/2}$, $\sigma_{Ea} = 0.3|f|Ri^{-1/2}$, and $\sigma_{CI} = (-f(f - U_y))^{1/2}$, respectively.

2a) with three increasing levels of eddy viscosity and diffusivity, $\kappa = 10^{-6}$, $\kappa = 10^{-4}$, and $\kappa = 10^{-2}$ m²/s restricts the wavenumber range over which there are nonzero growth rates until eventual elimination (Fig. 5). Recall that SI has canonical wavelength of 20-500 m in the global ocean (Dong et al., 2021a). The lowest wavenumbers (corresponding to SI with the largest wavelengths, or features with greatest spatial scales) are most resilient to damping by viscous-diffusive effects; which include viscous redistribution of TKE through the water column (Fig. 5) as well as TKE removal by dissipation ($\epsilon$). Dimensionalization of $l^*$ and $k^*$ gives an unstable wavelength of $\lambda_y = 2\pi H/l^*$, showing that adding a moderate ($\kappa = 10^{-4}$ m²/s) amount of diffusivity reduces the meaningful range of SI from near-inviscid wavelengths of 25-625m to wavelengths of 125-625m.

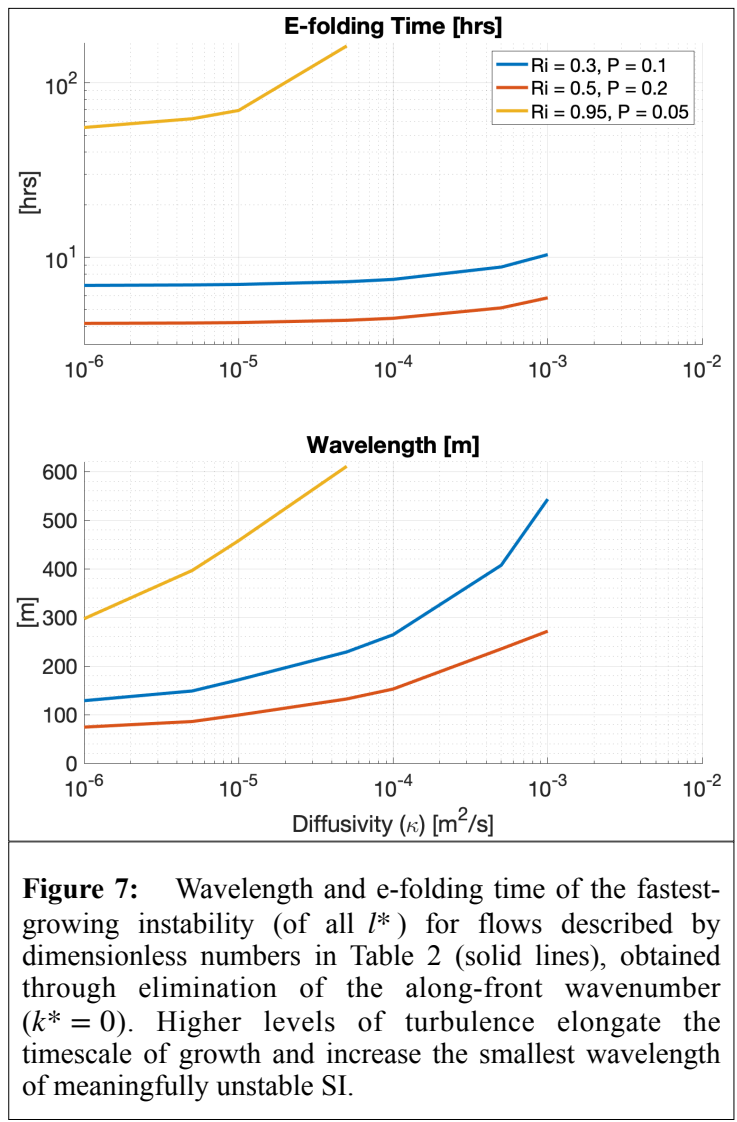

**Figure 7:** Wavelength and e-folding time of the fastest-growing instability (of all $l^*$) for flows described by dimensionless numbers in Table 2 (solid lines), obtained through elimination of the along-front wavenumber ($k^* = 0$). Higher levels of turbulence elongate the timescale of growth and increase the smallest wavelength of meaningfully unstable SI.

High levels of diffusivity (appropriate for a surface boundary layer experiencing shear-convective forcing) eliminate the symmetric mode altogether. When unstable SI modes are not eliminated the growth rate is reduced such that the e-folding time, $\tau = (\sigma^* U_z)^{-1}$, is elongated from 7 hrs to 10 hrs in the high North Atlantic mixed layer or 4 hrs to 6 hrs in the Antarctic Circumpolar Current mixed layer. Critically, the growth of an instability depends on the accumulation of turbulent kinetic energy ($K_t > 0$) produced via shear and/or buoyancy production. In the case of an oceanic flow with preexisting turbulence, there is a competition between production and removal by dissipation ($\epsilon$) by microscale features. This scenario is illustrated in Fig. 6, where secondary effects (vertical flux driven by pressure and viscosity) are net zero in our idealized setup. The influence of diffusivity ($\kappa$) on the maximum growth rate ($\sigma$) and associated wavelength ($\lambda_y$) of the fastest-growing SI wavenumber for the flows in Table 2 is given in Fig. 7; and is repeated for a range of Prandtl ratio and Richardson numbers (Fig. 8). Instability is eliminated for diffusivities $\kappa \geq 5 \times 10^{-5}$ m²/s, beginning with large Richardson numbers (low shear:stratification) and small Prandtl ratios (large stratification:rotation).

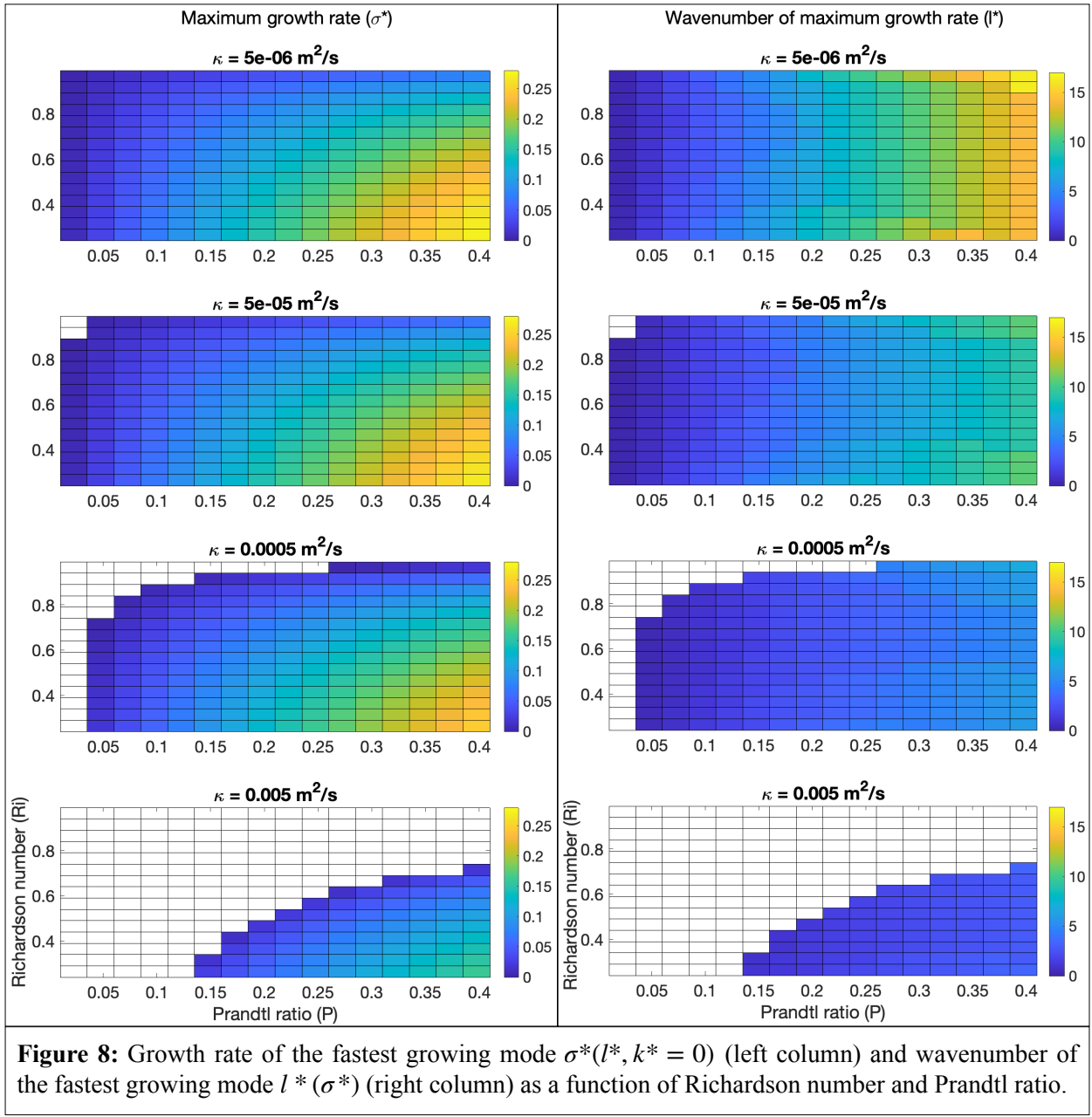

**Figure 8:** Growth rate of the fastest growing mode $\sigma^*(l^*, k^* = 0)$ (left column) and wavenumber of the fastest growing mode $l^*(\sigma^*)$ (right column) as a function of Richardson number and Prandtl ratio.

**Discussion**

From this heavily-idealized numerical instability analysis, we find that viscous-diffusive effects from preexisting turbulence can reduce the growth rate and wavelength range of SI in the ocean, and should be considered when developing parameterizations of SI-related mixing in circulation models. In situations with preexisting turbulence from competing instabilities---or SI itself---as commonly found at frontal jets, the growth of SI may not be as widespread as its identification based on submesoscale gradients (e.g. $qf < 0$) would suggest. Our results are summarized by a map of the stability boundary (Fig. 9) as a function of Richardson number, Prandtl ratio, and level of preexisting turbulence mixing. We conjecture that flows with low stratification (large $P$ or small $Ri$) are most resilient to viscous-diffusive effects of background turbulence, with turbulence eliminating SI entirely as $Ri \to 1$.

Viscous and diffusive effects restrict wavenumber range $(k, l)$ of the symmetric mode and reduce growth rate ($\sigma$) where SI persists in spite of damping by turbulence, corresponding to an elongation in e-folding time by $\mathcal{O}(1\text{-}10)$ hrs and limitation of spatial scale from $\mathcal{O}(10\text{-}100)$ to $\mathcal{O}(100)$ m. However the finding that turbulence reduces growth of frontal instabilities in the real ocean is unsurprising, given viscous/diffusive effects have long been employed to control numerical instabilities. These results provide one plausible explanation for the difficulty of observing SI in nature. Another possible explanation is that ageostrophic shear modifies total shear sufficient to alter stability in externally-forced regimes, resulting in an SI phenomenology[7] which does not match that formulated by the traditional Ertel potential vorticity criterion (Hoskins, 1974).

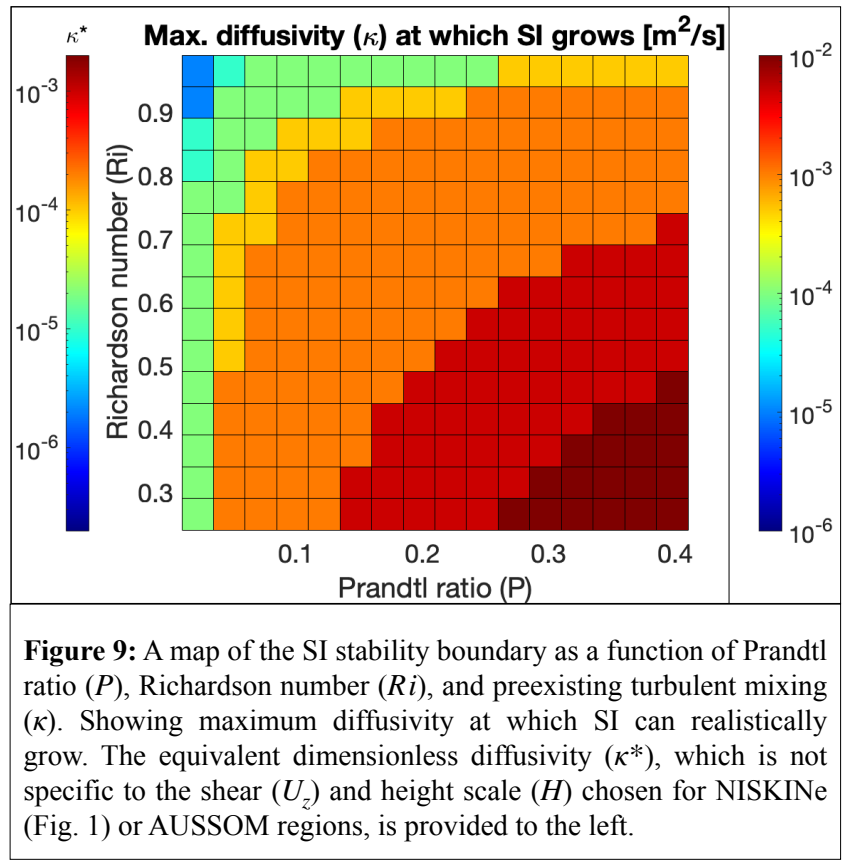

**Figure 9:** A map of the SI stability boundary as a function of Prandtl ratio ($P$), Richardson number ($Ri$), and preexisting turbulent mixing ($\kappa$). Showing maximum diffusivity at which SI can realistically grow. The equivalent dimensionless diffusivity ($\kappa^*$), which is not specific to the shear ($U_z$) and height scale ($H$) chosen for NISKINe (Fig. 1) or AUSSOM regions, is provided to the left.

This study investigated the effects of turbulence by imposing a uniform viscosity and diffusivity, $Pr = 1$. We varied the the magnitude of the viscosity and diffusivity via Reynolds number ($Re$), as well as the magnitude of stratification and shear (relative to each other and to planetary rotation) via the Prandtl ratio ($P$) and Richardson number ($Ri$). While it serves as an inquiry into the effects of mixing on instability in a stratified rotating system, our results are by no means generalizable to all cases of symmetric instability --- a point well-demonstrated by the diverse impacts of viscosity and diffusivity on parallel shear instability (Thorpe et al., 2013; Li et al.,

---

[7] Furthermore, SI hybridized with stratified shear instability in an ageostrophic flow would likely result in a physical signature different from pure SI.

2015; Smyth & Carpenter, 2019; Hughes et al., 2021). We take a moment to discuss these relevant studies, as the literature is considerably more extensive for this non-rotating setup and describes some nuances which potentially have rotating analogues. Thorpe et al. (2013) studied the stability of parallel shear flow on its neutral curve (the disturbance wavenumber and Richardson number at which an inviscid non-diffusive flow for which the flow is barely unstable), implemented a vertically varying viscosity and diffusivity. They conjectured that viscosity always stabilizes a shear flow (through energy removal, as in this paper) and diffusivity always destabilizes a shear flow (by diminishing stratification and allowing faster growth rates). They disproved their conjecture, finding that the combined effects of viscosity and diffusivity can either stabilize or destabilize a marginally stable flow, and there is no obvious rule to predict this behavior.

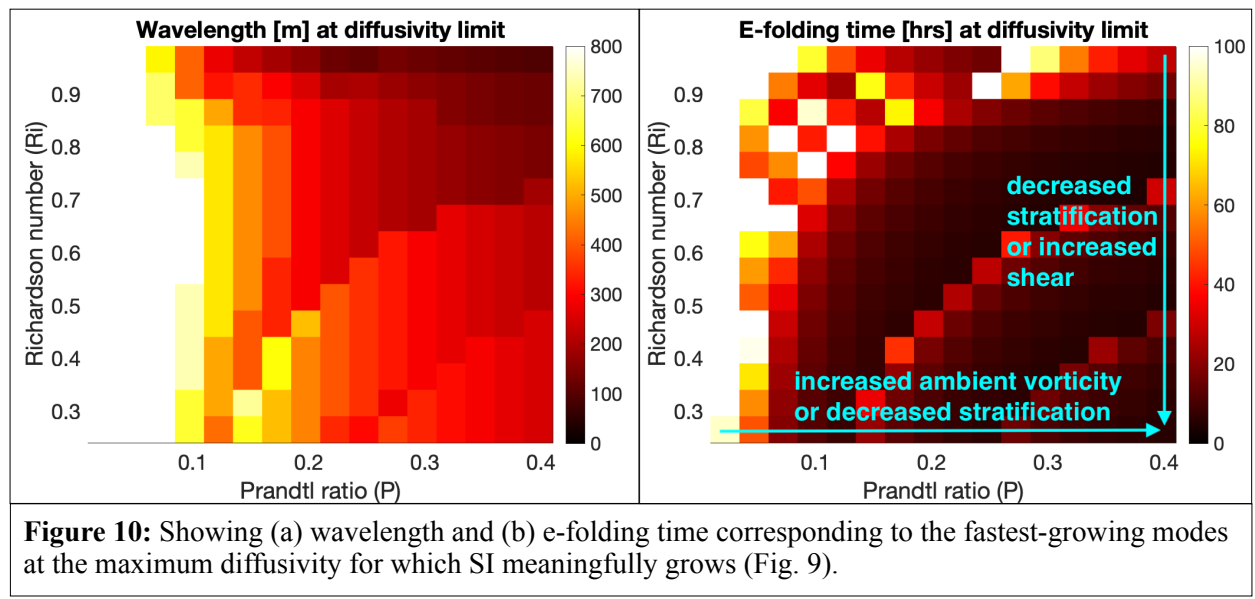

**Figure 10:** Showing (a) wavelength and (b) e-folding time corresponding to the fastest-growing modes at the maximum diffusivity for which SI meaningfully grows (Fig. 9).

We consider whether the resilience of instabilities persisting under high levels of viscosity and diffusivity (Fig. 9) is attributable to their spatial scale or growth rate in a given flow (Fig. 10). It is helpful, for comparison purposes, to know that our dimensionless numbers span $2.4 < \Gamma < 204.1$ (see Fig. 2b, Fig. 9a of Wienkers et al., 2021), such that the growth rate is not expected to vary strongly (e.g., logarithmically) with $P$ for a given $Ri$. Turbulence increases the wavelength of the fastest-growing unstable mode *for a given flow* (Fig. 7) which agrees with the dependence of dissipation (Eq. 4g) on the squared wavenumber $\tilde{k}^2 = l^2$; lower wavenumber and higher viscosity can have a compensatory effect on growth rate. In other words, larger wavelengths imply greater separation between the scale of the growing perturbation and the Kolmogorov scale, $\nu^{3/4}\epsilon^{-1/4}$. In agreement with Fig. 7, short wavelengths (dark red) are absent from increasing levels of viscosity and diffusivity. Large wavelengths themselves, however, are not the principal factor associated with the resilience of SI to intense mixing (Fig. 10a). For the surviving SI flows $(Ri, P)$ at a given level of viscosity and diffusivity, the growth rate increases with decreasing $Ri$ and increasing $P$ (Fig. 10b). As in Fig. 7, elongated timescales (medium red) are found at the outer $(Ri, P)$ limit for each higher level of turbulence, supporting the notion that viscosity and diffusivity decrease the growth rate (or increase the e-folding time) for the fastest-growing mode of a given flow.

Noting (Eq. 4c-d), SI extracts energy by advecting vertical and lateral shear ($U_z$, $U_y$, $V_z$, $V_y$) of the along-front current (Smyth & Carpenter, 2019). It is unsurprising that the most resilient symmetric instabilities persist in cases of low stratification:rotation (similar to Thorpe et al., 2013) and high shear:stratification. Thorpe et al. (2013) remark that wavenumber $k$ controls the impact of viscosity on the growth rate of shear instability; but canonical (inviscid, non-diffusive) SI is distinct from shear instability in that it has no specific wavelength and a growth rate that depends on flow characteristics, $\sigma_{SI}* = P\sqrt{1 - Ri}$. This is in contrast to shear instability for which the wavelength ($\lambda = c_1 h$) and growth rate ($\sigma = c_2 u_0/h$) of the fastest-growing mode are convolved with the geometry of the shear layer $h$ and thus with one another (Smyth & Carpenter, 2019).

Li et al. (2015) extended the work of Thorpe et al. (2013) by moving the flow setup away from the stability boundary (e.g. dropping $Ri$ from 0.25, in analogue to this study); they found the stabilizing or destabilizing effects of viscous diffusion reverse as baseline stability is decreased. At very small Richardson numbers (e.g. $Ri < 0.13$), the turbulent diffusion of vorticity dispersed gradients in the enstrophy profile, weakening the instability; whereas the opposite was true for $Ri > 0.13$. In other words, the interplay between turbulence and a growing instability is not only a function of energy but also enstrophy. Hughes et al. (2021) implemented a constant $\kappa = \nu = 3 \times 10^{-4} m^2 s^{-1}$ but varied the shear and stratification profiles to demonstrated the impact of the instability's proximity to the sea surface on its growth. SI is intrinsically tied to boundaries by forcing; and so the behavior of more complicated shear and stratification profiles, including those containing the buoyancy structure of competing instabilities (such as Kelvin-Helmholz billows), must not be overlooked.

We studied the net effect of a constant viscosity and diffusivity but did not investigate their individual roles; much could be done by varying the viscosity in space, magnitude, and magnitude relative to diffusivity --- all aspects of submesoscale instability which may have practical implications for subgrid-scale mixing parameterizations. More comprehensive numerical instability analyses (including variation of dimensionless numbers) are needed to further contextualize this result in the physical ocean, as well as other growth characteristics. One such characteristic is the ocean structural evolution resulting from mixing-arrested SI. We speculate that a symmetrically unstable front persists until mixing from competing instabilities and/or preexisting turbulence weakens the frontal gradients to a state of stability; or be modified by surface forcing until its characteristics ($P > P_{t=0}$, $Ri < Ri_{t=0}$, $K$) were *sufficiently unstable* to produce SI with a nonzero growth rate. Furthermore, most attention has been paid to the modal growth instabilities at stratified unstable fronts (Stone, 1970; Stamper & Taylor, 2017), but its transient effects may also be significant. Zemskova et al. (2020) examined transient (non-modal) growth of instabilities in the hydrostatic Eady problem, finding that associated growth rates are up to two orders of magnitude larger than those of the baroclinic and symmetric modes, likely acting as an non-negligible restratifying agent despite their asymptotic decay. We highlight the open tasks of (a) validating these findings with observations of SI in the ocean, (2) determining the resilience of hybrid types of SI and those with centrifugal influence to varying levels of turbulence, and (3) working towards a more dynamically-realistic SI parameterization which incorporates factors such as ageostrophic shear, preexisting turbulence, and topographic shearing.

**Acknowledgements**

We acknowledge invaluable feedback from Alexis Kaminski, Eric D'Asaro, and William (Bill) Smyth; and thank Bill for publishing a brilliant open-access textbook on the subject of instability. Ferris's effort was supported under ONR Northern Ocean Rapid Surface Evolution (NORSE) DRI award N000142112701 and the Applied Physics Laboratory - University of Washington Science & Engineering Enrichment & Development Fellowship. Data production for the NISKINe DRI was supported by ONR award N000141812386 (PI Harper Simmons). AUSSOM was supported by the OCE Division of the National Science Foundation, award 1558639 (PIs Louis St. Laurent and Sophia Merrifield). ROMS models were used courtesy of Simmons and John Pender, for which computational resources were provided by Research Computing Systems at University of Alaska Fairbanks. We thank Justin Shapiro for operating gliders *Husker* and *Starbuck*.

**Appendix A**

We obtain the normal mode equations for a perturbation in rotating, stratified, viscous, diffusive flow:

$$ik\hat{u} + il\hat{v} + \hat{w}_z = 0 \qquad \text{(X1a)}$$
$$(\sigma + ikU)\hat{u} + U_z\hat{w} = -ik\hat{\pi} + f\hat{v} + \nu\nabla^2\hat{u} \qquad \text{(X1b)}$$
$$(\sigma + ikU)\hat{v} = -il\hat{\pi} - f\hat{u} + \nu\nabla^2\hat{v} \qquad \text{(X1c)}$$
$$(\sigma + ikU)\hat{w} = -\hat{\pi}_z + \hat{b} + \nu\nabla^2\hat{w} \qquad \text{(X1d)}$$
$$(\sigma + ikU)\hat{b} + B_y\hat{v} + B_z\hat{w} = \kappa\nabla^2\hat{b} \qquad \text{(X1e)}$$

Here $\sigma \leftrightarrow \partial/\partial t$, $ik \leftrightarrow \partial/\partial x$, $il \leftrightarrow \partial/\partial y$, and $\nabla^2 \leftrightarrow \partial^2/\partial z^2 - \tilde{k}^2$ and $\tilde{k}^2 = k^2 + l^2$. $\hat{u}$ and $\hat{v}$ are perturbation velocity components, $\hat{b}$ is the buoyancy fluctuation, $\hat{\pi}$ is the pressure fluctuation, $k$ is the wavenumber aligned with the direction of the flow (along-front direction), $l$ is orthogonal to it (across-front direction), $U$ is flow velocity, and $B$ is background buoyancy. Our goal is to reduce the number of equations (through the elimination of $\hat{\pi}$ and replacement of $\hat{u}, \hat{v}$ with vorticity) before solving as an eigenvalue problem.

To obtain the vorticity equation we cross differentiate (Eq. X1b-c):

$$(\sigma + ikU)\hat{q} - ilU_z\hat{w} = f\hat{w}_z + \nu \nabla^2 \hat{q} \tag{X2}$$

To obtain the vertical momentum equation, we differentiate momentum equations (Eq. X1b) by $\partial/\partial x$, (Eq. X1c) by $\partial/\partial y$, and (Eq. X1d) by $\partial/\partial z$. We note $-\pi_{zz} = -\pi \tilde{k}^2 - \nabla^2 \hat{\pi}$, and have:

$$(\sigma + ikU)\frac{\partial \hat{u}}{\partial x} + ikU_z \hat{w} = k^2 \hat{\pi} + ikf\hat{v} + ik\nu \nabla^2 \hat{u} \tag{X3a}$$

$$(\sigma + ikU)\frac{\partial \hat{v}}{\partial y} = l^2 \hat{\pi} - ilf\hat{u} + il\nu \nabla^2 \hat{v} \tag{X3b}$$

$$(\sigma + ikU)\frac{\partial \hat{w}}{\partial z} + ikU_z w = -\pi \tilde{k}^2 - \nabla^2 \hat{\pi} + \hat{b}_z + \nu \frac{\partial}{\partial z}\nabla^2 \hat{w} \tag{X3c}$$

Defining perturbation vorticity $\hat{q} = ik\hat{v} - il\hat{u}$, noting continuity (Eq. X1a), and differentiating the sum of (Eq. X3a-c) by $\partial/\partial z$ (neglecting inflections[8] of $U$) isolates $\hat{\pi}$ for subsequent elimination.

$$\nabla^2 \hat{\pi}_z = -2ikU_z \hat{w}_z + \hat{b}_{zz} + f\hat{q}_z \tag{X4}$$

Differentiating (Eq. X1d) by $\nabla^2$:

$$(\sigma + ikU)\nabla^2 \hat{w} + 2ikU_z \hat{w}_z = -\nabla^2 \hat{\pi}_z + \nabla^2 \hat{b} + \nu \nabla^4 \hat{w} \tag{X5}$$

and substituting in (Eq. X4) into (Eq. X5) obtains the vertical momentum equation:

$$(\sigma + ikU)\nabla^2 \hat{w} = -\tilde{k}^2 b - fq_z + \nu \nabla^4 \hat{w} \tag{X6}$$

To obtain the buoyancy equation, we begin with (Eq. X1e). The definition of $q$ and of continuity (Eq. X1a) are rearranged to obtain

$$\hat{v} = -\hat{q}\frac{ik}{\tilde{k}^2} + \hat{w}_z \frac{il}{\tilde{k}^2}$$

This expression is substituted into (Eq. X1e) to eliminate $\hat{v}$, giving:

$$(\sigma + ikU)\hat{b} - B_y \hat{q}\frac{ik}{\tilde{k}^2} + B_y \hat{w}_z \frac{il}{\tilde{k}^2} + B_z \hat{w} = \kappa \nabla^2 \hat{b} \tag{X7}$$

(Eq. X2, Eq. X6, Eq. X7) indicate the reduced system of equations, and recasting as an eigenvalue problem gives

---

[8] See footnote 3. Neglect of velocity inflections is common for inquires into rotating stratified systems (e.g. Stamper & Taylor [2017]) but is worthy of future questioning, as they are known to (de)stabilize non-rotating shear flows.

$$\sigma \begin{pmatrix} I & 0 & 0 \\ 0 & \nabla^2 & 0 \\ 0 & 0 & I \end{pmatrix} \begin{pmatrix} \hat{q} \\ \hat{w} \\ \hat{b} \end{pmatrix} = \begin{pmatrix} -ikU + \nu\nabla^2 & ilU_z + fD^{(1w)} & 0 \\ -fD^{(1q)} & -ikU\nabla^2 + \nu\nabla^4 & -\tilde{k}^2 \\ \frac{ik}{\tilde{k}^2}B_y & -\frac{il}{\tilde{k}^2}B_y D^{(1w)} - B_z & -ikU + \kappa\nabla^2 \end{pmatrix} \begin{pmatrix} \hat{q} \\ \hat{w} \\ \hat{b} \end{pmatrix}$$

(X8)

where $D^{()}$ are vertical derivative operators (see footnote 4). After solving the eigenvalue problem (Eq. X8) for growth rate $\sigma$ and eigenfunctions ($\hat{q}$, $\hat{w}$, $\hat{b}$), the remaining perturbation quantities are obtained from algebraic solution of the normal mode equations. (Eq. X1a) is rearranged to become $\hat{u} = \hat{w}_z i/k - \hat{v}l/k$ and vorticity ($q$) is rearranged to become:

$$\hat{u} = \hat{v}k/l + \hat{q}i/l \quad \text{(X9a)}$$

such that equating these expressions for $\hat{u}$ allows us to solve for the across-front velocity:

$$\hat{v} = i(\hat{w}_z l - \hat{q}k)/\tilde{k}^2 \quad \text{(X9b)}$$

The pressure is obtained from (Eq. X1c) and vertical displacement is obtained from the time derivative of vertical velocity:

$$\hat{\pi} = -(\sigma\hat{v} + ikU\hat{v} + f\hat{u})/(il) + \nu\nabla^2\hat{v}/(il) \quad \text{(X9c)}$$
$$\hat{\eta} = w/(\sigma + ikU) \quad \text{(X9d)}$$

Together, Eq. X8 and Eq. X9 become Eq. 2 in the main text of the paper.

**References**

Bachman, S.D., Fox-Kemper, B., Taylor, J.R. and Thomas, L.N., 2017. Parameterization of frontal symmetric instabilities. I: Theory for resolved fronts. Ocean Modelling, 109, 72-95.

Bachman, S.D. and Taylor, J.R., 2014. Modelling of partially-resolved oceanic symmetric instability. Ocean Modelling, 82, 15-27.

Boccaletti, G., Ferrari, R. and Fox-Kemper, B., 2007. Mixed layer instabilities and restratification. Journal of Physical Oceanography, 37(9), 2228-2250.

Callies, J., Flierl, G., Ferrari, R. and Fox-Kemper, B., 2016. The role of mixed-layer instabilities in submesoscale turbulence. Journal of Fluid Mechanics, 788, 5-41.

Dong, J., Fox-Kemper, B., Zhang, H., & Dong, C., 2021a. The Scale and Activity of Symmetric Instability Estimated from a Global Submesoscale-Permitting Ocean Model. Journal of Physical Oceanography, 51(5), 1655-1670.

Dong, J., Fox-Kemper, B., Zhu, J. and Dong, C., 2021b. Application of symmetric instability parameterization in the Coastal and Regional Ocean Community Model (CROCO). Journal of Advances in Modeling Earth Systems, 13(3), p.e2020MS002302.

Ferris, L., 2022: Across-scale energy transfer in the Southern Ocean. Ph.D. dissertation, Dept. of Physical Sciences, Virginia Institute of Marine Science, William and Mary, 190 pp.

Ferris, L., Gong, D., Clayson, C.A., Merrifield, S., Shroyer, E.L., Smith, M. and Laurent, L.S., 2022a. Shear turbulence in the high-wind Southern Ocean using direct measurements. Journal of Physical Oceanography, 52(10), 2325-2341.

Ferris, L., Gong, D., Merrifield, S. and Laurent, L.S., 2022b. Contamination of Finescale Strain Estimates of Turbulent Kinetic Energy Dissipation by Frontal Physics. Journal of Atmospheric and Oceanic Technology, 39(5), 619-640.

Hoskins, B. J., 1974. The role of potential vorticity in symmetric stability and instability. Q. J. R. Meteorol. Soc., 100, 480-482.

Hughes, K.G., Moum, J.N., Shroyer, E.L. and Smyth, W.D., 2021. Stratified shear instabilities in diurnal warm layers. Journal of Physical Oceanography, 51(8), 2583-2598.

Kaminski, A. K. and Smyth, W.D., 2019. Stratified shear instability in a field of pre-existing turbulence. Journal of Fluid Mechanics, 862, 639-658.

Li, L., Smyth, W.D. and Thorpe, S.A., 2015. Destabilization of a stratified shear layer by ambient turbulence. Journal of Fluid Mechanics, 771, 1-15.

Lian, Q., Smyth, W., Liu, Z., 2020. Numerical Computation of Instabilities and Internal Waves from In Situ Measurements via the Viscous Taylor–Goldstein Problem. Journal of Atmospheric and Oceanic Technology, 37, 759-776.

Liu, Z., Thorpe, S.A. and Smyth, W.D., 2012. Instability and hydraulics of turbulent stratified shear flows. Journal of Fluid Mechanics, 695, 235-256.

Rees, T., Monahan, A., 2014. A General Numerical Method for Analyzing the Linear Stability of Stratified Parallel Shear Flows. Journal of Atmospheric and Oceanic Technology, 31, 2795-2808.

Skyllingstad, E. D., Duncombe, J., Samelson, R.M., 2017: Baroclinic frontal instabilities and turbulent mixing in the surface boundary layer. Part II: Forced simulations. J. Phys. Oceanogr., 47, 2429-2454.

Skyllingstad, E. D., Samelson, R. M., 2020: Instability processes in simulated finite-width ocean fronts. J. Phys. Oceanogr., 50, 2781-2796.

Smyth, W.D., 2008. Instabilities of a baroclinic, double diffusive frontal zone. J. Phys. Oceanogr., 38, 840-861.

Smyth, W.D., 2020. Marginal instability and the efficiency of ocean mixing, 50, 2141-2150.

Smyth, W.D., Burchard, H. and Umlauf, L., 2012. Baroclinic interleaving instability: A second-moment closure approach. Journal of physical oceanography, 42(5), 764-784.

Smyth, W.D. and J. R. Carpenter, 2019: Instability in Geophysical Flows. Cambridge University Press, 327 pp.

Smyth, W.D. and Ruddick, B., 2010. Effects of ambient turbulence on interleaving at a baroclinic front. Journal of physical oceanography, 40(4), 685-712.

Stamper, M.A. and Taylor, J.R., 2017. The transition from symmetric to baroclinic instability in the Eady model. Ocean Dynamics, 67(1), 65-80.

Stone, P. H. (1970). On non-geostrophic baroclinic stability: Part II. Journal of Atmospheric Sciences, 27(5), 721-726.

Thomas, L. N., Taylor, J. R., Ferrari, R., Joyce, T. M., 2013. Symmetric instability in the Gulf Stream. Deep Sea Res. Part II Top. Stud., 91, 96-110.

Thorpe, S.A., Smyth, W.D. and Li, L., 2013. The effect of small viscosity and diffusivity on the marginal stability of stably stratified shear flows. Journal of Fluid Mechanics, 731, 461-476.

Whalen, C.B., Drushka, K. and Gaube, P., 2022, December. Global Scale Variability of Submesoscale Frontal Dynamics. In AGU Fall Meeting Abstracts (Vol. 2022,. OS12A-03).

Whalen, C. B., et al., Global distribution and governing dynamics of submesoscale density fronts. (in prep)

Wienkers, A.F., Thomas, L.N. and Taylor, J.R., 2021. The influence of front strength on the development and equilibration of symmetric instability. Part 1. Growth and saturation. Journal of Fluid Mechanics, 926.

Yankovsky, E., Legg, S. and Hallberg, R., 2021. Parameterization of submesoscale symmetric instability in dense flows along topography. Journal of Advances in Modeling Earth Systems, p.e2020MS002264.

Zemskova, V. E., Passaggia, P. Y., White, B. L., 2020. Transient energy growth in the ageostrophic Eady model. Journal of Fluid Mechanics, 885.